\documentstyle{article}

\newcommand{\Rr}{{\rm I\!R}}

\newcommand{\half}{\frac{1}{2}}

\newcommand{\C}{{\mathchoice {\setbox0=\hbox{$\displaystyle\rm C$}\hbox{\hbox
to0pt{\kern0.4\wd0\vrule  height0.9\ht0\hss}\box0}}
{\setbox0=\hbox{$\textstyle\rm C$}\hbox{\hbox
to0pt{\kern0.4\wd0\vrule  height0.9\ht0\hss}\box0}}
{\setbox0=\hbox{$\scriptstyle\rm C$}\hbox{\hbox
to0pt{\kern0.4\wd0\vrule  height0.9\ht0\hss}\box0}}
{\setbox0=\hbox{$\scriptscriptstyle\rm C$}\hbox{\hbox
to0pt{\kern0.4\wd0\vrule  height0.9\ht0\hss}\box0}}}} 

\newcommand{\Z}{{{\mathchoice  {\hbox{$\textstyle  Z\kern-0.4em  Z$}}
{\hbox{$\textstyle  Z\kern-0.4em  Z$}}
{\hbox{$\scriptstyle  Z\kern-0.3em  Z$}}
{\hbox{$\scriptscriptstyle  Z\kern-0.2em  Z$}}}}}

\newcommand{\Q}{{\mathchoice   {\setbox0=\hbox{$\displaystyle\rm
Q$}\hbox{\raise
0.15\ht0\hbox  to0pt{\kern0.4\wd0\vrule  height0.8\ht0\hss}\box0}}
{\setbox0=\hbox{$\textstyle\rm  Q$}\hbox{\raise
0.15\ht0\hbox  to0pt{\kern0.4\wd0\vrule  height0.8\ht0\hss}\box0}}
{\setbox0=\hbox{$\scriptstyle\rm  Q$}\hbox{\raise
0.15\ht0\hbox  to0pt{\kern0.4\wd0\vrule  height0.7\ht0\hss}\box0}}
{\setbox0=\hbox{$\scriptscriptstyle\rm  Q$}\hbox{\raise
0.15\ht0\hbox  to0pt{\kern0.4\wd0\vrule  height0.7\ht0\hss}\box0}}}}

\newcommand{\qed}{\ \ \rule{1ex}{1ex}}

\newtheorem{theo}{Theorem}[subsection]
\newtheorem{prop}[theo]{Proposition}

\newtheorem{lem}[theo]{Lemma}
\newtheorem{cor}[theo]{Corollary}

\begin{document}

\begin{center}{\bf{\Large Addendum: Level spacings for integrable quantum maps in
genus zero 
 }}\footnote{
Partially supported by  NSF grant
\#DMS-9703775.}\\ 
\medskip

  Steve Zelditch\\
\medskip

{Johns Hopkins University, Baltimore, Maryland  21218}\\\medskip

April 29, 1998

\end{center}
\addtolength{\baselineskip}{1pt}

In this note we continue our study [Z] of 
 the pair correlation functions (PCF's)
$$\rho_2^n(f) = \frac{1}{n^2}\sum_{\ell} \hat{f}(\frac{\ell}{n}) |Tr U_n^{\ell}|^2|
= \frac{1}{n^2}\sum_{\ell} \hat{f}(\frac{\ell}{n}) |\sum_{ k =1}^n e^{i \ell n
[\alpha\phi(\frac{k}{n}) + \beta \frac{k}{n}]}|^2 $$ of completely integrable quantum
maps over $\C P^1$. To be specific, the quantum
maps are assumed to have  the form $U_{n, \alpha, \beta} = e^{i n (\alpha
\phi(\hat{I}) + \beta \hat{I})}$ where $\hat{I}$ is an action operator (i.e. an
angular momentum operator) with eigenvalues $\frac{k}{n}$ ($k = -n, \dots, n$),
acting on the quantum Hilbert space ${\cal H}_n$ of nth degree spherical harmonics at
Planck constant $\frac{1}{n}$.  Also $\phi$ is a smooth function satisfying $\phi''
\not= 0$ on $[-1,1].$  Our main result was  \begin{theo}[Z] Let $n_m = [m (\log
m)^4]$.  Then for almost all $(\alpha, \beta)$ (in the Lebesgue sense),
$\rho_{2,\alpha, \beta}^{n_m} \rightarrow \rho_2^{POISSON}$ as $m \rightarrow
\infty$. \end{theo}

Our aim in this addendum is to strengthen this result to  almost everywhere
convergence to  Poisson along the entire sequence of Planck constants. The price
we pay is that  the results apply not to the individual $\rho_2^n$'s but to
 the average
\begin{equation}  \bar{\rho}_{2, \alpha, \beta}^{N} : = \frac{1}{N} \sum_{n=1}^N
\rho^{n}_{2, \alpha, \beta}.\end{equation}
Here we change the notation from $\rho_2^N$ in [Z] to $\rho_2^n$ so that $N$ is
reserved for the cumulative PCF $\bar{\rho}_2^N$ up to level $N$. 

\begin{theo} Suppose that $\phi(x)$ is a polynomial satisfying $\phi'' \not= 0$
on $[-1,1]$. Then,
for almost all $(\alpha, \beta)$ we have: $$ \bar{\rho}_{2, \alpha, \beta}^{N}
\rightarrow \rho_2^{POISSON}.$$ \end{theo}

This addendum was    motivated by a comparison of the results of [Z] with those of
 Rudnick-Sarnak [R.S] on the PCF of fractional parts of polynomials. 
Independently, both [R.S] and [Z] established mean square convergence to Poisson
of their respective PCF's.  However, [R.S] went on to prove a.e. convergence. Their
technique was first to prove that the local
PCF's $\rho_{2,\alpha, \beta}^{n_m}$ tend to Poisson almost everywhere along a sparse
subsequence $\{n_m\}$ of Planck constants, and then to show 
that for $n \in [n_m, n_{m+1}]$ the oscillation $\rho_2^n - \rho_2^{n_m}$ was
relatively small and hence the full sequence converged to Poisson.  
 This latter step seemed
(and still seems) 
intractable in the  quantum maps situation [Z].  The main difference is
that the local spectra in [R.S] increase with $n$ whereas for quantum maps [Z] they
change in rather uncontrollable ways. However we can re-establish a parallel to their
situation by focussing on  
 the mean PCF's  $ \bar{\rho}_{2, \alpha,
\beta}^{N}$  rather than the individual $\rho_2^{n}$'s.  Our spectra then
increase with $N$ and there is much less oscillation between Planck constants.

As in [R.S], the proof of this last step is based on the use of
Weyl estimates of  exponential sums and seems limited to polynomial phases. 
In addition to the  Weyl method, it also uses some considerations from
the measure theory of continued fractions.

\section{Preliminary results on $ \bar{\rho}_{2, \alpha,
\beta}^{N}$ }

Up until the last step, the analysis of $ \bar{\rho}_{2, \alpha,
\beta}^{N}$ is analogous to the analysis of $ \rho_{2, \alpha,
\beta}^{N}$ in [Z].  As in [Z, Theorem (5.1.1)] we have:
\begin{theo} Let $\hat{H}_{\alpha, \beta} = \alpha \phi(\hat{I}) + \beta
\hat{I}$ where $|\phi''|\geq C_o > 0$ on $[-1,1].$  Let $\bar{\rho}_{2, \alpha,
\beta}^N$ be as above.  Then for any $f$ with supp$\hat{f}$ compact:
$$\int_{-T}^T  \int_{-T}^T |\bar{\rho}_{2, \alpha, \beta}^N (f) -
\rho_2^{POISSON}(f)|^2 d \alpha d\beta = O(\frac{(\log N)^2}{N}).$$ \end{theo}

\begin{cor} Let $N_m = [m (\log m)^4]$.  Then for almost all $(\alpha, \beta)$
in the Lebesgue sense, 
$$\lim_{m \rightarrow \infty} \bar{\rho}_{2, \alpha, \beta}^{N_m} (f) =
\rho_2^{POISSON}(f).$$ \end{cor}

To fill in the gaps in the sparse susequence $\{N_m\}$, consider $\bar{\rho}^M_{2,
\alpha, \beta}$ for
 $N_m < M
< N_{m+1}.$  Obviously,  \begin{equation} \bar{\rho}_{2, \alpha, \beta}^{M}(f) -
\bar{\rho}_{2, \alpha, \beta}^{N_m}(f) = \frac{N_m - M}{M} \bar{\rho}_{2, \alpha,
\beta}^{N_m}(f) + \frac{1}{M}\sum_{n = N_m}^M \rho_{2, \alpha, \beta}^{n} (f).
\end{equation}

We have $M - N_m << (N_{m+1} - N_m) \sim
(m + 1) (\log(m+1)^4) - m (\log m)^4  << (\log m)^4.$  So in the first sum
$\frac{N_m - M}{M} << m^{-1 + \epsilon}.$  In the second we have $O( (\log m)^4)$
terms. Under the assumption 
supp$\hat{f} \subset [-1,1]$ the trivial bound $\rho_2^n(f) << n$ already gives
\begin{equation} \frac{N_m - M}{M} \bar{\rho}_2^{N_m}(f) + 
\frac{1}{M}\sum_{n = N_m}^M \rho_{2, \alpha, \beta}^{n} (f) << (M - N_m) << (\log
m)^4 . \end{equation}
So we just need a tiny improvement on the trivial bound to prove that these
terms tend to zero.  In the following section we will prove that for almost all
$(\alpha, \beta)$, $\rho_{2, \alpha, \beta}^{n} (f) \leq C(\alpha, \beta) n^{1 -
\frac{2}{K} + \epsilon}$ where $K = 2^{k-1}$ with $k$ the degree of $\phi$.   From
this it also follows by standard density arguments that $\bar{\rho}^N_{2,
\alpha,\beta} [a, b] \rightarrow \rho^{POISSON}_2 [a,b]$ for all intervals $[a,b]$. 
We refer to [R.S] for the details of the density argument.

\section{The Main Lemma}
 
The purpose of this section is to prove:
 \begin{lem} Suppose that $\phi$ is a polynomial of degree $k$ satisfying 
 the hypotheses: (i) $|\phi''|>0$ and (ii) $|\alpha \phi' + \beta| > 0$ on
$[-1,1]$, .  Then for any $\hat{f} \in C_o(\Rr)$ and almost all $(\alpha,\beta)$, we
have:   $n^2 \; \bar{\rho}_{2, \alpha, \beta}^{(n)} (f) \leq C(\alpha, \beta) n^{1 -
\frac{2}{K} + \epsilon}$, where $K = 2^{k-1}$. \end{lem}

Recall that the local PCF's have the form
$$\rho_{2, \alpha, \beta}^n = \sum_{\ell \in \Z} \hat{f}(\frac{\ell}{n}) |\sum_{k =
1}^n  e(\alpha n \ell [\phi(\frac{k}{n}) + \beta \frac{k}{n}])|^2.$$
Since $\hat{f}$ is compactly supported,  the  $\ell$-sum  runs over
an interval of integers of the form $[- C n, C n]$ for some $C > 0.$  For simplicity
of notation, and with no loss of generality, we will assume the sum over $\ell$
runs over the interval $[-n,n].$
Throughout we use the  notation $e(x) = e^{2\pi i x}.$

\subsection{The quadratic case}

The case of quadratic polynomials is more elementary than that of polynomials
of general degree and we can prove our main result
without analysing continued fraction convergents to $\alpha$.  Hence we begin by
discussing this case.
The relevant exponential sum is 
$$|\sum_{k = 1}^n 
e(\alpha n \ell [\phi(\frac{k}{n}) + \beta \frac{k}{n}])|^2 = 
\sum_{h = -n}^n \sum_{x = 1}^{2n} e(\ell h (\alpha \frac{x}{n} + \beta)).$$
For $f$ with supp$\hat{f}$ in $[-1,1]$ we have
$$n^2 \; \rho_{2, \alpha, \beta}^n(f) << |\sum_{|\ell| \leq n}\sum_{h = -n}^n \sum_{x =
1}^{2n} e(\ell h (\alpha \frac{x}{n} + \beta))|.$$

The following estimate is weaker than that claimed in the Main Lemma but is
sufficient for the proof of the theorem. 
 
\begin{lem} Let $\alpha$ be a diophantine number satisfying $|\alpha -
\frac{a}{q}| \geq \frac{K(\alpha)}{q^{2 + \epsilon}}$ for any rational number
$\frac{a}{q}.$  Then
for all $\beta$,  $\rho_{2, \alpha, \beta}^n(f) << n^{\half + \epsilon}.$ \end{lem}

\noindent{\bf Proof}:

 We begin with the standard estimate (e.g. [K, Lemma 1])
$$|\sum_{x = 1}^{2n} e(\ell h
(\alpha \frac{x}{n} + \beta))| = |\sum_{x = 1}^{2n} e(\ell h
(\alpha \frac{x}{n} ))|  \leq \min (2n, \frac{1}{2 ||\ell h \frac{\alpha}{n}||})$$
where $|| \cdot||$ denotes the distance to the nearest integer. This gives 
$$n^2 \rho_{2, \alpha, \beta}^n(f) << \sum_{\ell \leq n}\sum_{h = -n}^n \min (2n,
\frac{1}{2 ||\ell h \frac{\alpha}{n}||}). $$
The variable $x = h \ell$ runs over $[-n^2, n^2]$; when $x\not= 0$, the  multiplicity
$c_x = \# \{(h,\ell): h \ell = x\}$ is well-known to have order $n^{\epsilon}$ (e.g
[V, Lemma 2.5]).  Then there are $2 n$ terms where $h \ell = 0$, each contributing
$n$ to the sum.  Hence, 

 \begin{equation} n^2 \rho_{2, \alpha, \beta}^n(f) << n^2 +  n^{\epsilon} \sum_{x =
-n^2}^{n^2} \min (2n, \frac{1}{2 ||x \frac{\alpha}{n}||}). \end{equation}

At this point we are close to the well-known estimate ( e.g.
Korobov [K, Lemma 14])
$$\sum_{x = 1}^{Q} \min (P, \frac{1}{||\alpha x + \beta||}) <<
(1 + \frac{Q}{q})(P + q \log P)$$
where     $\alpha = \frac{a}{q} + \frac{\theta}{q^2}$ with
$|\theta| < 1$ and with $(a,q) = 1.$  In our situation $Q = n^2, P = n$, giving
$( 1 + \frac{n^2}{q})(n + q \log n)$, but
the estimate does not apply because our `$\alpha$' is $\frac{\alpha}{n}$;
 the  rational approximation $\frac{a}{qn}$ to   $\frac{\alpha}{n}$ has a
remainder  of only $\frac{1}{n q^2}$ rather than $\frac{1}{(nq)^2}.$  This
complicates the argument and worsens the resulting estimate.  

 Since we do not know the continued fraction expansion of $\frac{\alpha}{n}$, we
use the rational approximation
$\frac{\alpha}{n} = \frac{a}{qn} + \frac{\theta}{n q^2}$. It is not necessary
that $(a, n) = 1$ so we rewrite $\frac{a}{qn} = \frac{a'}{q n'}$ with 
$(a', n') = 1$ (hence $(a', n' q) = 1$).  Then
$$\frac{\alpha}{n} = \frac{a'}{n' q} + \frac{\theta}{nq^2},\;\;\;\;\;(a', n') = 1,
\;\;\;\;\;|\theta| < 1.$$

Now break up $[-n^2, n^2]$ into blocks of length $n' q$.  There
are at most $2 [\frac{n^2}{ n' q}] + 1$ such blocks.  
Hence  
\begin{equation} n^{\epsilon} \sum_{x = -n^2}^{n^2} \min (2n, \frac{1}{2 ||x
\frac{\alpha}{n}||}) << n^{\epsilon} 
\sum_{y = 0}^{[\frac{n^2}{n' q}] + 1}\sum_{x = 1}^{n' q} \min (2n,
\frac{1}{2 ||(x + y q n' )\frac{\alpha}{n} ||}).\end{equation}

The above rational approximation brings
$$\frac{\alpha x}{n} + y q n'\frac{\alpha}{n} = \frac{a ' x }{n ' q} + 
\frac{x \theta}{n q^2} + y a' + \frac{y  n' \theta}{n q}.$$
Hence
$$||\frac{\alpha x}{n} + y q n'\frac{\alpha}{n}|| =|| \frac{a ' x }{n ' q} + 
\frac{x \theta}{n q^2} + \beta||$$
where $\beta = \{\frac{y  n' \theta}{n q}\}$.  Write $\beta = \frac{b(y)}{n' q}
+ \frac{\theta_1}{n'q}$ with $b(y) \in \Z$ and with $|\theta_1| < 1$. Since $|x|
\leq n' q$ we have $$||\frac{a' x + b(y)}{n' q}|| = ||x \frac{\alpha}{n}
 + y q n'\frac{\alpha}{n} - \frac{x \theta}{n q^2} - \frac{\theta_1}{n' q}||   \leq 
||x \frac{\alpha}{n}
 + y q n'\frac{\alpha}{n}|| 
+ \frac{1}{n' q} + \frac{n'}{n q}.$$
The remainder $\frac{n'}{nq}$ is much larger than occurs in the
standard argument and since it is possible that $n' = n$ we can only be sure that
the remainder is $O(\frac{1}{q}).$

Therefore we are only sure that our sum is 
 $$<<   n^{\epsilon} 
\sum_{y = 0}^{[\frac{n^2}{n' q}] + 1}\sum_{x = 1}^{n' q} \min (2n,
\frac{1}{2 ||\frac{a ' x + b(y)}{n ' q} + O(\frac{1}{q}) ||}).$$ 

Since $(a', n' q) = 1$, the numbers $a'x + b(y)$ run thru a complete residue system
modulo $n' q$ as  $x$ runs thru $1, \dots n'
q$.  Hence, the $x$-sum is
independent of $a', b(y)$ and we may rewrite it as
$$  <<  (\frac{ n^{2 + \epsilon}}{n' q} + 1)  \sum_{2
\leq x \leq n' q- 1} \min(2n,  \frac{2}{||\frac{x}{n' q} + O(\frac{1}{q})||}).$$ 
The distance $||\frac{x}{n' q} + O(\frac{1}{q})||$ can be less than $\frac{1}{n}$
over the range of terms $x \in [0, C n']$ and $x \in [n'q - C n', n' q]$ where
$C$ is the implicit constant in $O(\frac{1}{q}).$  For these we must take $n$
in the minimum.  Since there are $O(n)$ such terms in the $x$-sum, their contribution
to the entire sum is $<< n^{2 + \epsilon} \frac{n^2}{n' q}.$

For the remaining terms we use that $\min(2n, \frac{2}{||\frac{ x}{n' q}||})$
is an even function of $x$ to put the $x$-sum in the form
$$   \sum_{C n' \leq x \leq \frac{q n'}{2}}
\min(2n,  \frac{2}{||\frac{x}{n' q} + O(\frac{1}{q})||}).$$ 
The minimum is now surely attained by $ \frac{2}{||\frac{x}{n' q} +
O(\frac{1}{q})||}$ and since it stays in the left half of the interval we have
$$\frac{1}{||\frac{x}{n' q} + O(\frac{1}{q})||} = \frac{1}{\frac{x}{n' q} +
O(\frac{1}{q})}.$$
Therefore
$$   \sum_{C n' \leq x \leq \frac{q n'}{2}}
\min(2n,  \frac{2}{||\frac{x}{n' q} + O(\frac{1}{q})||}) <<
n' q \sum_{C n' \leq x \leq \frac{q n'}{2}} \frac{1}{x - O(n')}
<< n' q \log (n' q).$$ 
The whole $x$-sum is therefore $<< (\frac{n^{2 + \epsilon}}{n' q} + 1) [n^2 + n 'q
\log (n' q)].$

In sum,  we have
$$n^{\epsilon}\sum_{x = -n^2}^{n^2} \min (2n, \frac{1}{ ||x
\frac{\alpha}{n}||}) << (\frac{n^{2 + \epsilon}}{n' q} + 1) [n^2 +  q n' \log (n'
q)].$$  Hence
$$ \rho_{2, \alpha, \beta}^n << 1 + (\frac{n^{ \epsilon}}{n' q} + n^{-2}) [n^2 +  q
n' \log (n' q)].$$  The first parenthetical term  is  of size $n^{1 + \epsilon}/q$
when $n' = n$ while  the trivial bound was $n$. 
It is at this point that we must   restrict to diophantine numbers satisfying
$|\alpha - \frac{a}{q} | \geq \frac{K(\alpha)}{q^{2 + \epsilon}}$ for all
rational $\frac{p}{q}.$ By Dirichlet's box principle there exists $q \leq n^r$ and a 
rational $\frac{a}{q}$ with $(a,q) = 1$ such that
$|\alpha - \frac{a}{q}| \leq \frac{1}{q n^r}.$ It follows that
$q > n^{r - \epsilon}.$ Substituting into our estimate, we get 
$$ \rho_{2, \alpha, \beta}^n << 1 + ( \frac{n^{- r + \epsilon}}{n' } + n^{-2})  [n^{2
} + n^r n'  \log (n) ] << n^{\epsilon} ((a,n) n^{1-r} +
\frac{1}{(a,n)} n^{-1 + r}.$$

Since $1 \leq (a,n) \leq n$ the final estimate is
$$<<  n^{\epsilon}( n^{2 - r} + n^{-1 + r} ).$$
The terms balance when $r = \frac{3}{2}$ to give
$$\rho_{2, \alpha, \beta}^n(f) << n^{\half + \epsilon}.$$ \qed

\noindent{\bf Remark} In the next section we will see that there are rational
numbers $\frac{a}{q}$ satisfying the above requirements and also satisfying $(a, n)
\leq C(\alpha)  n^{\epsilon}.$  This  changes the final estimate to
$<<  n^{\epsilon}( n^{1 - r} + n^{-1 + r} )$ and gives
 $\rho_{2, \alpha, \beta}^n(f) << n^{ \epsilon}.$

\subsection{The general polynomial case}

Now let $\phi(x) = \alpha_o x^k + \alpha_1 x^{k-1} \dots + \alpha_k$ be a 
general polynomial.  We would like to estimate
$$\rho_2^n(f) = \frac{1}{n^2} \sum_{\ell \in \Z} \hat{f}(\frac{\ell}{n})
|\sum_{k=1}^n e(n \ell \phi(\frac{x}{n}))|^2. $$
As in the classical Weyl inequality (cf. [V, Lemma 2.4]) we will 
estimate $|\sum_{k=1}^n e(n \ell \phi(\frac{x}{n}))|^2$ by squaring and  differencing
repeatedly until we reach the linear case.  
Let $\Delta_j$ be the jth iterate of the forward difference operator, so that
$$\begin{array}{l} \Delta_1 \phi (x ; h) = \phi(x + h) - \phi(x) \\
\Delta_{j+1} \phi (x; h_1, \dots, h_{j+1}) = \Delta_1 (\Delta_{j} \phi(x; h_1,
\dots,h_j; h_{j+1})). \end{array}$$

We recall (cf. [V, Lemma 2.3]):

\begin{lem} We have
$$ |\sum_{x = 1}^n e(f(x))|^{2^j} \leq (2 n)^{2^j - j - 1}
\sum_{|h_1| < n} \cdots \sum_{|h_j| < n} [\sum_{x \in I_j} 
e(\Delta_j f(x; h_1, \dots, h_j))]$$
where the intervals $I_j = I_j(h_1, \dots, h_j)$ satisfy $I_1 \subset [1, n]$,
$I_j \subset I_{j-1}.$  \end{lem}

Now let 
$$T(\phi; n, \ell) = \sum_{x = 1}^n e(n \ell \phi(\frac{x}{n}))$$
with $\phi(x) = \alpha_o x^k + \dots + \alpha_o$ and put $K = 2^{k-1}.$  Apply the
previous lemma with $j = k-1$  to get:
$$ |T(\phi; n, \ell)|^{K} << n^{K - k} \times$$
$$\sum_{h_1} \cdots \sum_{h_{k-1}}
 \sum_{x \in I_{k-1}} e(h_1 \dots h_{k-1} \ell p_{k-1}(x;
h_1, \dots, h_{k-1}; n, \ell)).$$
Here, the sum runs over $h_j$ with $|h_j| \leq n$ and
$$p_{k-1}(x; h_1, \dots, h_{k-1}; n) =  k! n^{-k + 1}
\alpha_o (x + \half h_1 +
\dots + \half h_{k-1}) + (k-1)! n^{-k + 2}\alpha_1.$$
This is just as in the standard Weyl estimate ([V][D, \S 3]) except for the
powers of $n$ in the coefficients of $p_{k-1}.$

Then write 
\begin{equation} \rho_2^n(f) = \frac{1}{n} \sum_{\ell } \hat{f}(\frac{\ell}{n})
[\frac{1}{n} |T(\phi; n, \ell) |^2] << \frac{1}{n} \sum_{\ell \leq n} 
(\frac{1}{n} |T(\phi; n, \ell) |^2) \end{equation}
Since the $\ell$-sum is an average, we may apply Holder's inequality 
with exponent $\frac{K}{2}$ to get
\begin{equation} \rho_2^n(f) << [ \frac{1}{n} \sum_{\ell \leq n}  |\frac{1}{\sqrt{n}}
T(\phi; n, \ell) |^{K}]^{\frac{2}{K}} \end{equation}
Therefore
$$[\rho_2^n(f)]^{\frac{K}{2}} << n^{K - k} n^{-\frac{K}{2} - 1} \sum_{\ell \leq n}  
 \sum_{h_1} \cdots \sum_{h_{k-1}}
 \sum_{x \in I_{k-1}} e(h_1 \dots h_{k-1} \ell p_{k-1}(x;
h_1, \dots, h_{k-1}; n)).$$

There are $n^{k - 1}$ terms with $h_1 \dots h_{k-1}\ell = 0$, each contributing
$n$ to the $x$-sum.  So the contributions of such terms to the total sum is
$ O(n^{k})$, and we get
 \begin{equation} [\rho_2^n(f)]^{\frac{K}{2}} << n^{\frac{K}{2} - k - 1} [ n^{k}
 + \sum_{\ell \leq n}
\sum_{h, x}'  e(h_1 \dots h_{k-1} \ell p_{k-1}(x;
h_1, \dots, h_{k-1}; n))] \end{equation}
where the primed sum runs only over non-zero values of $h_1 \dots h_{k-1} \ell.$

As in the case with $k = 2$ above we sum over $x$ to get
\begin{equation} [\rho_2^n(f)]^{\frac{K}{2}} <<  n^{\frac{K}{2} - k - 1} [ n^{k} +
\sum_{\ell \leq n} \sum_{h}' min (n, \frac{1}{||k! h_1 \dots h_{k-1}\ell n^{-k +
1}\alpha||})]\end{equation} and then rewrite the variable $k! h_1 \dots h_{k-1}\ell$
as a new variable $x$ ranging over $[0, k! n^{k}].$ As before, the number  $c_x$ of
ways of representing $x\not= 0$ as  a product $k! h_1 \dots h_{k-1}\ell$  is
$O(n^{\epsilon})$ so  \begin{equation} [\rho_2^n(f)]^{\frac{K}{2}} <<  n^{\frac{K}{2}
- k - 1 + \epsilon} [ n^{k} +  \sum_{x \leq k! n^{k} }  min (n, \frac{1}{||x n^{-k +
1}\alpha||})].\end{equation}

We now repeat the steps of the quadratic case but with $\frac{\alpha}{n^{k -1}} $
replacing $\frac{\alpha}{n}.$  Thus, the rational approximation $\alpha = \frac{a}{q}
+  \frac{\theta}{q^2}$ gives the approximation 
$\frac{\alpha}{n^{k-1}} = \frac{a}{n^{k-1} q} + 
\frac{\theta}{n^{k-1} q^2}$ and hence requires us to break up the sum over
$[0, k! n^k]$ into blocks of size $n^{k-1} q/ (a, n^{k-1})$. Precisely the
same  argument (with $n'_k = \frac{n^{k-1}}{(a, n^{k-1})}$) then gives
$$\sum_{x \leq k! n^{k} }  min (n, \frac{1}{||x n^{-k +
1}\alpha||} ) << (\frac{ n^k}{q n'_k} + 1)(n^k + q n_k' \log (q n_k')).$$
Hence we get 
\begin{equation} [\rho_2^n(f)]^{\frac{K}{2}} << n^{\frac{K}{2} - k - 1 + \epsilon} [
n^{k} +  (\frac{  n^k}{q n'_k} + 1)(n^k + q n'_k \log(q n'_k))] <<
 n^{\frac{K}{2} - k - 1 + \epsilon}[n^{k} + \frac{  n^{2k}}{q n'_k} + q n_k'].
\end{equation}  
Recalling that $n'_k = \frac{n^{k - 1}}{(a, n^{k-1})}$ the last expression is
$$<< n^{\frac{K}{2} - 1 + \epsilon}[ 1  + \frac{n^k (a,n^{k-1})}{q n^{k - 1}} + 
\frac{q}{ n (a, n^{k-1})}].$$  Thus, 
\begin{equation} [\rho_2^n(f)] << n^{1 - \frac{2}{K} + \epsilon}[ 1  + \frac{n
(a,n^{k-1})}{q } +  \frac{q}{ n (a, n^{k-1})}]^{\frac{2}{K}}
\end{equation}
The exponent of the right side will be less than one if and only if the 
exponent of $[ 1  + \frac{n
(a,n^{k-1})}{q } +  \frac{q}{ n (a, n^{k-1})}]$ is less than one.  Thus we
are in very much the same situation as in the quadratic case (although the
resulting exponent will be increasingly bad as $K \rightarrow \infty$). 
However,   the  estimate $(a, n) \leq n$ used in the quadratic case  does
not generalize well to higher degree:   In higher degree, the estimate $(a, n^{k-1})
\leq n^{k-1}$ leads to $r = \frac{k+1}{2}$ and an exponent larger than one. 
Therefore we need to choose a rational approximation  satisfying $(a,q) = 1$ 
and $|\alpha - \frac{a}{q}| <
\frac{1}{q^2}$ and with low value of $(a, n^{k-1})$. 
The natural   candidates for such numbers are the continued fraction
convergents $\frac{p_m}{q_m} = [a_o, a_1, \dots, a_m]$ to $\alpha = [a_o, a_1,
\dots].$  Therefore we need to study the behaviour of
\begin{equation} f_n(\alpha): = \min\{\frac{n
(p_m(\alpha) ,n^{k-1})}{q_m(\alpha) } +  \frac{q_m(\alpha)}{ n (p_m(\alpha),
n^{k-1})}\}. \end{equation}
Since $\frac{p_m}{q_m} = \alpha + O(\frac{1}{q_m^2})$ we can (and will) replace the
$q_m$ in this definition by $p_m$
Since it is presumably hard to arrange for $(p_m(\alpha),
n^{k-1})$ to be large, we will require that $p_m(\alpha) \in [n^{r - \epsilon},
n^r]$ for some exponent $r$ to be determined later.  Before proceeding let us
recall how the index $m$ is related to $n, r$.

\begin{prop} For any $r, \epsilon > 0$, any $M \in {\bf N}$ and almost any $\alpha
\in \Rr$, there exists $n_o \in {\bf N}$ with the following property: for $n \geq n_o$
there exist at least $M$ consecutive convergents $p_{m-M}(\alpha), p_{m-M +
1}(\alpha), \dots, p_m \in [ n^{r - \epsilon}, n^r]$ with $m \leq C(\alpha) \log n.$
\end{prop}

\noindent{\bf Proof}:
By a theorem of Khinchin and Levy [Kh], one knows that for almost all $\alpha $ the
convergents satisfy \begin{equation} \lim_{m \rightarrow \infty} q_m ^{\frac{1}{m}} 
 = \gamma,\;\;\;\;\;\;\;\gamma:= \frac{\pi ^2}{12 \log 12}.\end{equation}
 
The first claim 
 is equivalent to the statement that there exists $m$ such that, for $0 \leq j \leq
M,$  $$(r -\epsilon) \log n < \log p_{m-j} = m \log \gamma + o(m) < r \log n.$$
Evidently there exists $C(\alpha) > 0$ such that $ m \leq  r C(\alpha) \log n$,
proving the second claim.  
The first claim is states that for sufficiently large $n$, there are at least $k$
consecutive solutions $m$ of $$[\frac{(r - \epsilon)}{\gamma} + o(1)] \log n  \leq m
\leq   [\frac{r}{\gamma} + o(1)] \log n.$$
This is obvious since the  width of the interval equals $[\frac{\epsilon}{\gamma} +
o(1)] \log n$, which is  positive and unbounded. \qed

We then have:
 
\begin{prop} Fix $k, r, \epsilon > 0$.  Then for almost all $\alpha \in \Rr$ there
exists a convergent $\frac{p_m(\alpha)}{q_m(\alpha)}$ with $p_m(\alpha) \in [n^{r -
\epsilon}, n^r]$ and with  $(p_m(\alpha), n^{k-1})  \leq n^{ \epsilon}.$ \end{prop}

\noindent{\bf Proof}  By the previous proposition, for any $M > 0$, there are at
least $M$  consecutive $p_m$'s in $[n^{r- \epsilon}, n^r]$ for sufficiently large
$n$.  Our goal is to find one satisfying $(p_m(\alpha), n^{k-1})  \leq n^{1 +
\epsilon}.$

To this end we recall [Kh] that 
$$\left\{ \begin{array}{l} p_m = a_m p_{m-1} + p_{m-2} \\
q_m = a_m q_{m-1} + q_{m-2} \end{array} \right. $$
and hence that $p_m q_{m-1} - p_{m-1} q_m = \pm 1.$
It follows that $p_m(\alpha), p_{m- 1} (\alpha)$ are relatively prime. This
pattern continues in a sufficiently useful way. By a simple induction we find that
for $k < m$, 
\begin{equation} p_m q_{m-k} - p_{m-k} q_k = \pm E_{k - 1} (a_m, a_{m-1}, \dots,
a_{m-k + 1})\end{equation} where $E_{0} = 1, E_1(a_m) = a_m, E_2(a_m, a_{m-1}) = a_m
a_{m_1} + 1$ and where $$E_{k} (a_m, a_{m-1}, \dots, a_{m-k }) = a_{m-k} E_{k-1}(a_m,
a_{m-1}, \dots, a_{m-k + 1}) + E_{k-2}(a_m, a_{m-1}, \dots, a_{m-k + 2}).$$
Hence any common divisor of $p_{m}, p_{m-1}, p_{m-2}$ is a divisor of $a_m$, and
so on.

We now claim that for the $M$ consective $p_m$'s in $[n^{r-\epsilon}, n^r]$ we have:
\begin{equation} (p_{m-M}, n^{k-1}) (p_{m - M + 1}, n^{k-1}) \cdots (p_m, n^{k-1})
\leq n^{k - 1}\;\; \Pi_{j = 0}^M \Pi_{\ell = 1}^{M - j} E_{\ell} (a_{m - j}, a_{m-j -
1 }, \dots, a_{m-j - \ell + 1 })\end{equation}

The idea of the argument is that, were all the $p_{m - j}$'s  relatively
prime, then each $(p_{m - j}, n^{k-1})$ would contribute a distinct factor of
$n^{k-1}$ and hence the product would be  $\leq n^{k-1}.$  The  
$p_{m - j}$'s are of course not relatively prime but (15) gives an upper bound on
the greatest common divisors of each pair.

 Thus, let us start with $p_m$ and consider
the degree to which factors in $(p_m, n^{k-1})$ are replicated by the
lower $(p_{m - j}, n^{k-1})$'s.  Since $(p_m, p_{m-1}) = 1$ there is no
duplication of factors  due to the  nearest neighbor. Since $(p_m, p_{m-2}) | a_m$ the
greatest common factor  of $(p_{m-2}, n^{k-1}), (p_m, n^{k-1})$ is less than
$(a_m, n^{k-1})$ and hence less than $a_m.$ Similarly the greatest common factor
of $(p_{m-3}, n^{k-1}), (p_m, n^{k-1})$ is less than $E_2(a_m, a_{m-1}).$  In all,
the  product $(p_{m-M}, n^{k-1}) (p_{m - M + 1}, n^{k-1}) \cdots (p_m, n^{k-1})$
replicates factors of $(p_m, n^{k-1})$ by at most 
$ E_1(a_m) \dots E_{M} (a_m, a_{m-1}, \dots, a_{m- M + 1 }).$ 

Next, move on to
$(p_{m-1}, n^{k-1}).$  These  factors of $n^{k-1}$ can get duplicated in $(p_{m-3},
n^{k-1})$ and so on down to $(p_{m-k}, n^{k-1}).$ One gets a similar estimate as in
the first case but with the indices lowered by one. Proceeding down to
$(p_{m-M},n^{k-1})$ proves the claim.

To complete the proof of the proposition, we use another fact from the metric
theory of continued fractions [Kh, Theorem 30]: For almost any $\alpha \in \Rr$,
 there exists $C(\alpha)>0$ such that $a_m(\alpha) \leq C(\alpha) m^{1 + \epsilon}$.
By  Proposition (2.2.2), the relevant  values of $m$ are of order $\log (n).$
Therefore, for the $p_m, p_{m-1}, \dots, p_{m-M}$ under consideration we have 
$a_{m - j} << \log n.$  Since $E_{\ell}$ is a polynomial in the $a_{m-
j}$'s of degree $\ell$, we have
$$E_{\ell } (a_{m - j}, a_{m-j - 1}, \dots, a_{m- j - \ell + 1 }) << (\log
n)^{\ell }.$$ Therefore
\begin{equation} \Pi_{j = 0}^M \Pi_{\ell = 1}^{M - j} E_{\ell } (a_{m - j}, a_{m-j
- 1 }, \dots, a_{m-j - \ell + 1 }) << (\log n)^{M^3}. \end{equation}
It follows that 
\begin{equation}  \Pi_{j = 0}^M (p_{m - j}, n^{k-1}) \leq C(\alpha) n^{k-1} 
(\log n)^{M^3}. \end{equation}
Hence at least one factor must be $\leq C(\alpha)^{1/M} n^{\frac{k-1}{M}}
(\log n)^{M^2}.$ The proposition follows from the fact that $M$ can be
arbitrarily large.\qed

We now complete the proof of the lemma and of our main result.  We have proved
the existence of $(p_m, q_m)$ with all the necessary properties and such that
$q_m \in [n^{r - \epsilon}, n^r], (p_m, n^{k-1}) << n^{\epsilon}.$  It follows
that 
\begin{equation} \frac{n (p_m ,n^{k-1})}{q_m } +  \frac{q_m}{ n (p_m, n^{k-1})}
<< n^{1 + \epsilon - r} + n^{r - 1}. \end{equation}
The terms balance when $r = 1/2$ and give the power $n^{ \epsilon}.$
It follows from (12) that $\rho_2^n(f) << n^{1 - \frac{2}{K} + \epsilon}.$ \qed

\end{document}